MICHELE TUCCI

DEPARTMENT OF PUBLIC ECONOMICS
FACULTY OF ECONOMICS
UNIVERSITY OF ROME "LA SAPIENZA"

# OLIGOPOLISTIC COMPETITION IN AN EVOLUTIONARY ENVIRONMENT: A COMPUTER SIMULATION

*Abstract*

The following notes contain a computer simulation concerning effective competition in an evolutionary environment. The scope is to underline the existence of a side effect pertaining to the competitive processes: the tendency toward an excess of supply by producers which operate in a strongly competitive situation. A set of four oligopolistic firms will be employed in the formal reconstruction. The simulation will operate following the Edmond Malinvaud "short side" approach, as far as the price adjustment is concerned, and the sequential Hicksian "weeks" structure with regard of the temporal characterization. The content of the present paper ought to be considered as a development of the writing: Michele Tucci, *Evolution and Gravitation: a Computer Simulation of a Non-Walrasian Equilibrium Model*, published with the E-print Archives at arXiv.com (section: Computer Science, registration number: cs.CY/0209017). A mirror is provided by SISSA.IT at the following address: http://babbage.sissa.it/abs/cs.CY/0209017
In such a paper there can be found some preliminary considerations regarding the comparison between the evolutionary and the gravitational paradigms and the evaluation of approaches belonging to rival schools of economic thought.

**Information on the author**: MICHELE TUCCI, Department of Public Economics, Faculty of Economics, University of Rome "La Sapienza", Italy.
Email: tucci@dep.eco.uniroma1.it
Web page: http://dep.eco.uniroma1.it/~tucci/

# *OLIGOPOLISTIC COMPETITION IN AN EVOLUTIONARY ENVIRONMENT: A COMPUTER SIMULATION*
Michele Tucci

**Note for the reader**
The following argumentation ought to be regarded as a development of the writing: Michele Tucci, *Evolution and Gravitation: a Computer Simulation of a Non-Walrasian Equilibrium Model*, published with the E-print Archives at arXiv.com (section: Computer Science, registration number: cs.CY/0209017). A mirror is provided by SISSA.IT at the following address:
http://babbage.sissa.it/abs/cs.CY/0209017
Therefore, the reader who is interested in the methodological analysis of the theoretical tools employed in the following exposition – especially with regard to the comparison between the evolutionary and the gravitational approaches – should refer to the cited paper.

**1 - Effective competition in an evolutionary environment**
The evolution of the technologies employed within an economic environment for the scope of producing goods is a central topic in virtually every school of economic thought, even in those founded on the gravitational paradigm. When economies based on the functioning of markets are taken into consideration, competition has always been indicated as the acting force which compels each supplier to improve its own productive structure. It is a matter of survival: whether or not a firm may be able to meet the challenge it depends on the global nature of its structure. As in every evolutionary process, lack of critical information prevents any a priori calculation about the possibilities of a productive unit to satisfy the new requirements: the entrepreneurs will be obliged to act according to their own subjective estimates, i.e. their own "animal spirits". Then the selection will occur: some companies will survive, some not.
Competition is a deadly matter: it always implies losses for the individuals and for the communities. The firms which are unable to stand the pace will disappear. The capital resources that had been used to create them will be wasted end labor will be unemployed. Still, such destruction will bear a most desired effect: the improvement of the global productive structure. New technologies, new products and new ways to organize the economical activities will come into existence and the world will be changed in an irreversible way: evolution takes place.

The scope of the present paper is to underline a side effect of competitive processes which has been rarely treated in literature: the tendency toward an excess of supply by producers which operate in a strongly competitive environment. Such a phenomenon will be analyzed within the framework of an evolutionary process by the usage of a simple computer simulation.

The core of the Darwinian evolutionary process is based on the occurrence of two conditions: the appearance of the mutations and the selection of the fittest. During the first phase the *status quo* is altered by the coming into existence of new elements which bear the potentiality of irreversibly altering the previous context. In the second stage, competition takes place: each element will have to face the struggle for survival. At the end, the winners will establish a renewed environment which will stimulate the coming into existence of a new set of alterations, thus starting the evolutionary process once again. It should be noticed that in the above very brief description the terms "element", "alteration" and "environment" have not been further qualified, leaving the phenomena under examination undefined. Of course Darwin was interested in analyzing the evolution of the living beings, but the evolutionary approach in itself is quite a general way to use our mind in order to reach a degree of understanding about the world occurrences. It is a tool to gain knowledge, rather than a feature belonging exclusively to the realm of biology. Within the field of economics, probably the most famous example of the implementation of an evolutionary framework can be found in the Schumpeterian reconstruction of technological progress concerning the firm productive structures[1].

Within the traditional neoclassical framework competition tends to be represented as an archetypal "just in time". Thought, of course, the genesis of this same approach must have followed the inverse path: quite probably it was built on the Walrasian concept of static equilibrium. In such a world competition is simply a matter of calculation. Since every agent possesses perfect knowledge, each producer is able to determine the production cost of every competitor on the market. Only those whose production cost will adhere to the minimum will actually produce, while the others will remain idle. Therefore, in such a context competition does not belong to the realm of phenomena, but to the one of the mind: it consists in calculations, not actions. Still, the more we analyze such a Platonic utopia, the more we are tormented by an uncomfortable sense of unreality. In fact, everyday experience stresses the fierce factual nature of the competitive processes. As it has been already pointed out, it appears to be a matter of life and death, both for the individuals and the communities. Taking into the due consideration Poincaré conventionalist school of epistemology, it may be argued that competition could be defined by using more than one model, and of course different models need not to share the same point of view concerning the analyzed object. While the Walrasian approach is a construction that belongs to the gravitational realm, on the contrary in our daily life, to accomplice the task of grasping the events of the economic world, we tend to employ an evolutionary approach. Nevertheless it is quite possible to build models that analyze economic phenomena from the evolutionary point of view: such being the task that is pursued with the present notes. If time were not to exist, Walrasian perfection would be reachable but, alas! Time does exist.

---

[1] See: TUCCI M. (2000).

In an evolutionary environment competition shows a very specific pattern of behavior: in order to effectively compete with each other, commodities must be present on the market at the same time and in the same place, so that the consumer may be able to exert his choice. The winning commodities will be bought and their producers will be able to obtain the due remuneration. The losing goods will remain unsold on the shelves of the retailers, until they will be destroyed or, in the most fortunate case, moved to less pretentious markets where they may possibly be sold at strongly discounted prices. In both cases, for the involved producers the future appears to be rather gloomy… Therefore effective competition – i.e. competition within an evolutionary environment – requires that each supplier, whose "animal spirits" induce him to dare challenging the market, will be obliged to produce at least the minimum amount that is needed to obtain a sufficient visibility. Then it is a matter of winning or losing… Hence, the basic principle of effective competition can be stated: in order to compete, firstly it is necessary to produce: production comes first, while competition will be the second stage.

Let's now proceed to lay down the fundamental assumptions employed in the simulation which will be shown in the following paragraph. The general framework is made up by a single market where four oligopolistic firms operate. The time cycle will be structured in a sequence of Hicksian "weeks". Within each time interval the market will be regulated by the "short side" disequilibrium approach. Each firm is supposed to possess perfect knowledge about its own productive processes and the extension of his own market share, while every other piece of information will be regarded as an expectation and , therefore, as knowledge the reliability of which is a matter of subjective probability. For the sake of simplicity, the following simulation will not deal with data structured in a probabilistic form, thought of course such a limit can be removed in more detailed models. Moreover, a general principle is holding to be true: if a firm lowers its selling price, the corresponding market share will tend to increase; and vice versa. Finally, it will be supposed that the firms will show an aggressive behavior: whenever it is possible, they will try to increase their own market shares, while they will be reluctant to decrease them.

In the next paragraph we will examine the technical details of the simulation.

4## 2 - The simulation

Let's consider a single market whose supply side is made up by four oligopolistic firms. Each one of them will employ capital and labor in order to produce amounts of the considered commodity. At the beginning of the cycle, for the sake of simplicity it will be supposed that the market share of each producer will be equal to one fourth, having set the aggregate demand constant through time and equal to one. The scope of each oligopolistic firm is to maximize its market share, by subtracting portion of it from the other competitors. Since, as it has already been pointed out, technological progress will not be taken into consideration, the only way each firm is able to compete is by riding their "lucky bits" – i.e. taking advantage of any favorable pricing of the productive factors that may eventually occurs.

Let's examine the structure of the simulation in a more detailed way[2]. Oligopolistic analysis rarely employs productive structures with constant returns to scale, since such an assumption is more suitable for a perfect competition environment. Moreover, pricing policies of oligopolistic firms tent to be a rather complicated affair. Still, in order to simplify the simulation, in the present case both constant returns to scale and selling prices equal to production costs will be assumed. To avoid modeling the financial structure of the firms, each company will be assumed to possess a monetary "buffer", i.e. a sum of money that can be used to cover eventual losses throughout the whole cycle, so that no firm will need to go bankrupted. The production structure of each unit will be represented by the traditional Cobb-Douglas function with constant returns to scale:

(1)     $y = x_1^c \, x_2^d$

where $x_1$ specifies capital and $x_2$ labor. Moreover, it will be: $c + d = 1$. Each firm will display a different set of the exponential coefficients (*c, d*) among the following: (0.20, 0.80), (0.40, 0.60), (0.60, 0.40), (0.80, 0.20). At the beginning of each time interval the prices of capital and labor will be casually altered, using the standard C++ function *rand( )*. For the sake of simplicity, at the beginning of the cycle it will be set $p_1 = 0.50$, $p_2 = 0.50$, where the symbols refer respectively to the price of capital and labor. The price normalization will be set such as: $p_1 + p_2 = 1$. Since each company has his own unique productive structure, changes in the prices of capital and labor will advantage some, while damaging others. The favored ones will be able to act aggressively, by lowering the selling price and trying to increase their market shares. On the contrary, those whose production costs have been increased will be obliged to rise their selling prices and, as a consequence, to give up portions of their market shares.

---

[2] The following simulation has been coded and compiled by the author using Microsoft Visual C++ 6.0.

Let's see some of the details relative to the way the behavior of the firms has been modeled. As it has already been pointed out, constant returns to scale will be supposed and the selling price will be set equal to the production cost by every competitor. At the beginning of each time interval, using the new prices of capital and labor, each firm will be able to calculate the corresponding selling price by using the standard Cobb-Douglas expression:

(2) $\quad p_y\_new = 1 / [\, (c / p_1\_new)^c \, (d / p_2\_new)^d \,]$

where $p_y\_new, p_1\_new, p_2\_new$ indicate respectively the selling price and the prices of capital and labor for the new time interval. Each company will now be able to compare the actual selling price, i.e. $p_y\_new$, with the selling price of the previous time interval, i.e. $p_y\_old$. Three circumstances may occur:

(3)
- (I) $\quad p_y\_new < p_y\_old$
- (II) $\quad p_y\_new > p_y\_old$
- (III) $\quad p_y\_new = p_y\_old$

In the first case, the selling price has been lowered. Therefore, the firm in question is able to act aggressively in order to increase its market share. In accordance with the already stated principle of effective competition, the same does necessarily require the increase of the production level. In fact, in order to compete, you must first produce. Only when your product will be on the shelves of the resellers side by side with the outputs of the competitors, so that the buyer is able to choose which one to buy, there will exist effective competition and there will be possible to distinguish the winners from the losers. Since the examined companies do not have any information about the behavior of the competitors, the amount of such an increase is a matter of expectations. The more the firm augments its own production, the more new retailers can be provided with its own products, whose price has been reduced, the more it is probable to induce the consumers to select its own output rather than that of the rival producers. The other side of the dilemma is that if too much is produced, chances are that some of it will remain unsold, thus constituting a loss. The mathematical expressions which have been used to represent the behavior of each firm are the following:

if case (I) in expression (3) holds:
*y_new* = *y_old* [ 1 + *gamma_one* * *atan*( $p_y$_old – $p_y$_new ) / ($\pi$ / 2) ]

(4)  if case (II) in expression (3) holds:
*y_new* = *y_old* [ 1 + *gamma_two* * *atan*( $p_y$_old – $p_y$_new ) / ($\pi$ / 2) ]

if case (III) in expression (3) holds:
*y_new* = *y_old*

where *y_new* and *y_old* are respectively the production level of the present time interval and of the previous one, *atan( )* is the standard C++ arctangent function and *gamma_one*, *gamma_two* are parameters setting the reaction sensitivity. For the sake of simplicity, the last two quantities will assume the same values for each one of the four firms. During the whole cycle, in each time interval the global demand will be set equal to one. Market behavior will follow the traditional disequilibrium "short side" approach: whenever there is an excess of supply, the exceeding part will be freely disposed; if the global demand tops the supply, then a portion of it will remain unsatisfied. In order to simplify the model, throughout the considered cycle the actual market share of each firm will be set in proportion with the relative amount of the produced commodity.

Simulations have been carried out with a large set of the parameter values. Whenever the behavior of the firms is set in an aggressive mode, in virtually every occurrence at the end of the cycle there has been a global excess of supply, defined as the sum of the excesses of supply for each firm and for every time interval. Such result constitutes the thesis of the present argumentation: in an evolutionary environment oligopolistic competition is structurally linked with the presence of excesses of supply – i.e. quantities of commodities that need to be produced in order to compete, but cannot be sold. Such a feature constitutes a loss both for the individual consumer and the whole economic community, since in the long run it will necessarily lead to a price increase. Vice versa, when competition is associated with technological progress, the same loss is still there, but the historical evidence shows that it's largely compensated by the benefit of a more efficient productive structure in such a way that the resultant of the combined tendencies is definitely favorable.

The graphs 1 ~ 4 illustrate an example of the simulation that has been described. There has been considered a cycle of 30 time intervals. Graph 1 shows the values that have been randomly assigned by the function *rand( )* to the prices of capital and labor. Graphs 2 and 3 indicate respectively the production cost and the production amount for each firm. Finally, graph 4 specifies the excess of supply during each time interval of the cycle. In the case under examination, the global excess of supply for the whole cycle can be approximated to 1.5. Since market demand has been considered constant during each time cycle and equal to one, we can conclude that there has been a global excess of production equal to one and a half the size of the demand level.

    Finally, it should be noted that the present simulation model ought to be considered as a prototype: more detailed software structures are needed in order to analyze the actual economic phenomena.

**1 - Commodity prices**

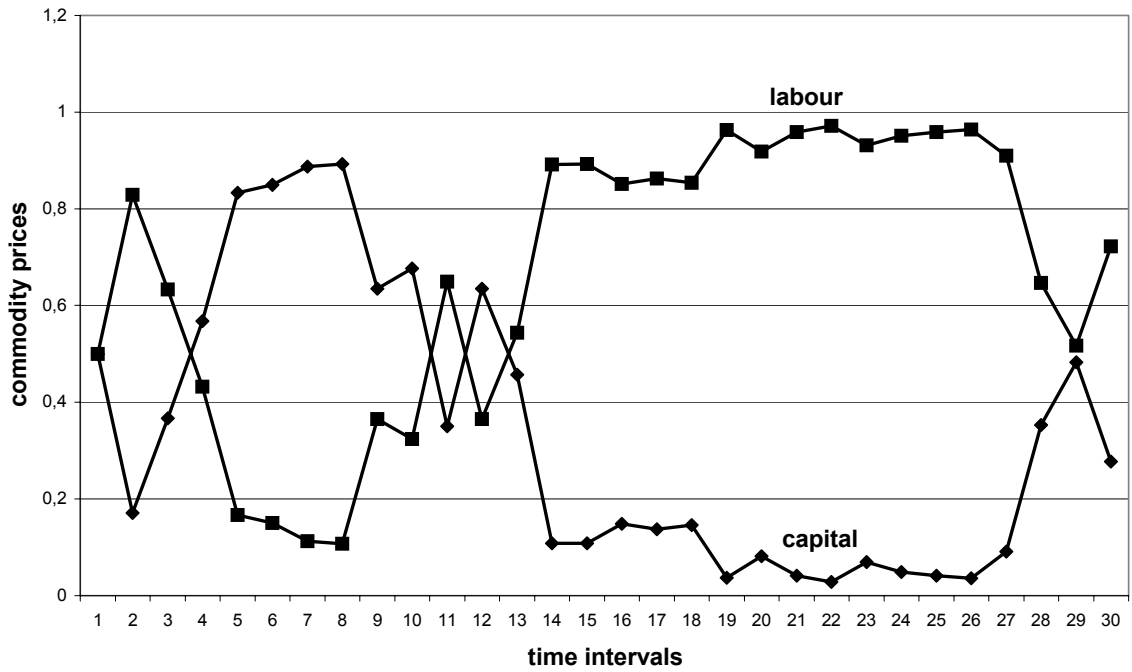

**2 - Production costs**

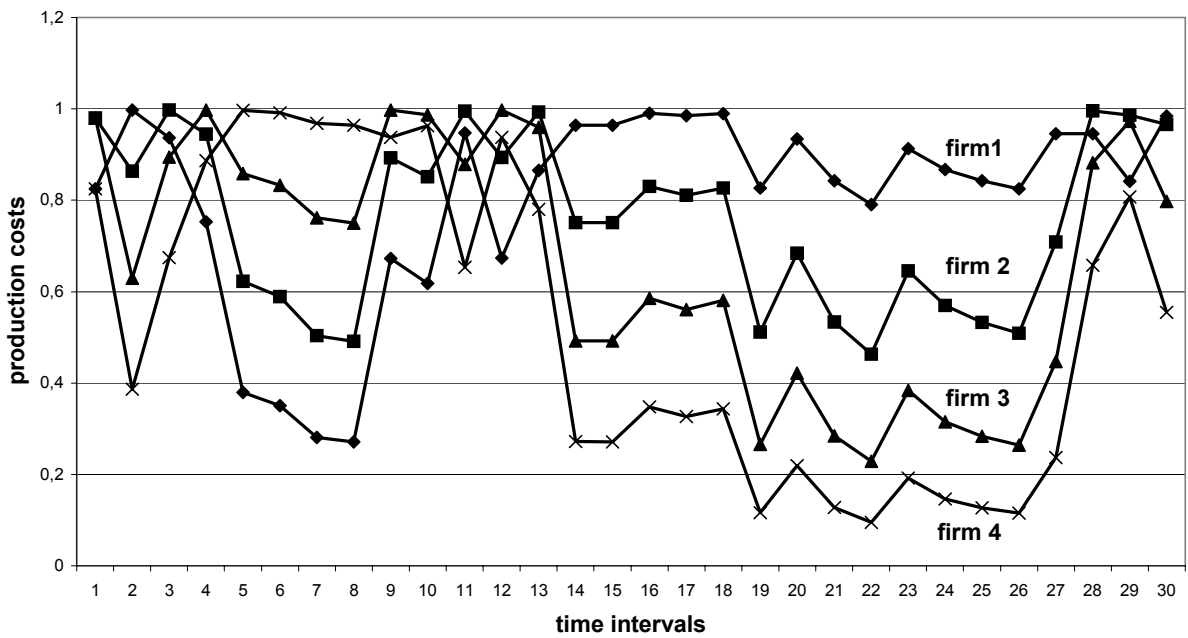

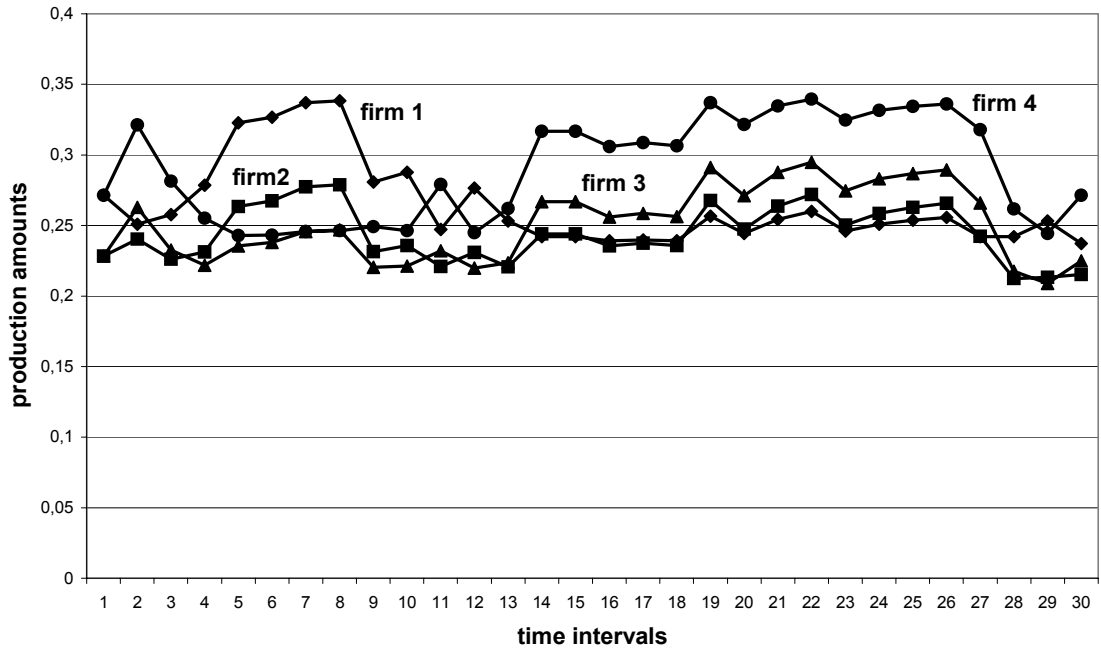

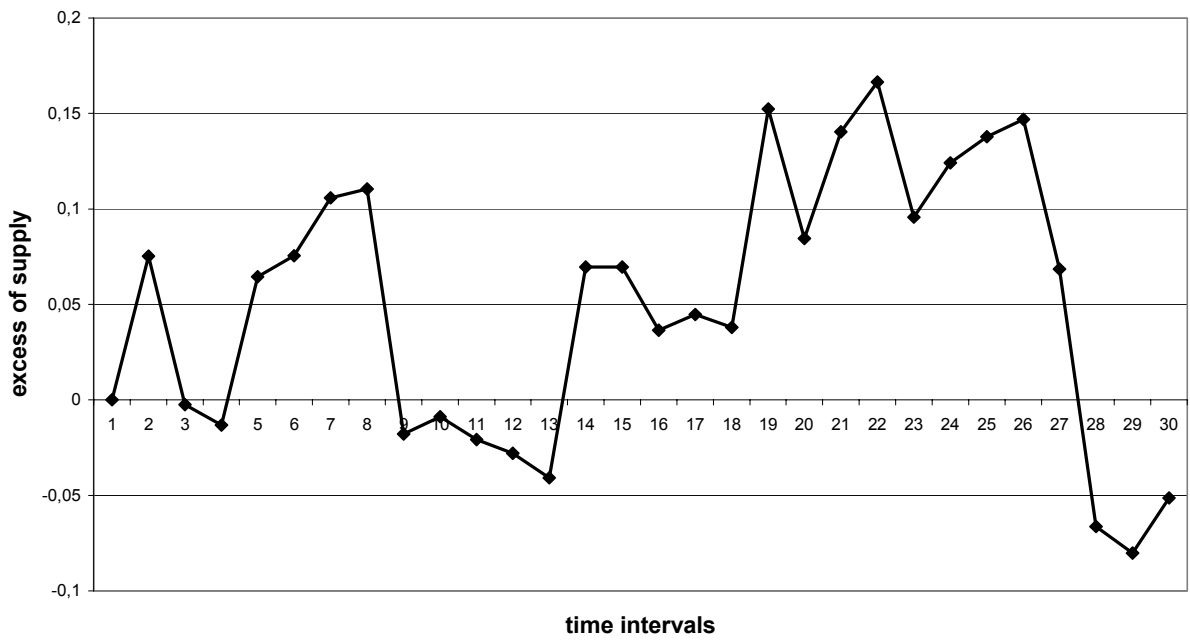

*Post Scriptum*

# Adversus Equilibrium

Equilibrium in economics is tantamount to General Economic Equilibrium theory, even if looser versions of such a concept have been related incorrectly to the classical school of economic thought. Such a point will be clarified later. For now, let's take into consideration GEE, which is the problematic aspect of the present status of the economic theory. To use an uncompromising expression, GEE is a "dead man walking": not really alive, but not gone yet. Such a situation is quite painful for the psyche of the economists, who therefore tend to avoid the issue. In their professional life, economists have no use for GEE. The great economical institutions of the world – governments, central banks, international associations, etc. – operate by using a collection of economic policy tools which don't have anything to do with GEE. In fact, the rationale of GEE would impose not to alter the supposedly "natural" behaviour of the markets, while the scope of the above quoted organizations is exactly to tamper with markets, hopefully for the best. Still, not very many economists would be willing to publicly declare that GEE belong to the history of economic thought and therefore should be substituted by more meaningful theoretical foundations. This is really a painful knot of the profession!

The philosophical idea underlying the concept of equilibrium used in GEE dates back to the Greek philosopher Parmenides, and it was developed more exhaustively by Plato. Still, it was not before the coming into existence of Newton's law that such a paradigm would assume a central role in the development of western sciences, mainly in the field of physics. The law of universal gravitation must have exerted an enormous influence on the minds of Newton's contemporaries: for the first time since ever mankind possessed a mathematical formula which would enclose a fragment of God's truth. There is no doubt that such state of mind strongly contributed to the emergence of the hubris which has characterized the development of the industrial revolution. Still, time goes by and starting with the end of the nineteenth century new ideas came to substitute Newton's divine certainties: Einstein's relativity theory firstly, then the quantistic approach would draw a rather different image of reality. Nowadays, physical universe seems to us a more mysterious and uncertain realm than it would have appeared to Newton's followers: once again, Eden is lost forever. Of course, in our daily life apples still fall from their trees! The gravitational paradigm is still applicable to the large majority of the physical phenomena that interfere with our daily lives. Therefore, the point is the following: is economics one of the realms of experience which can be ruled by the gravitational paradigm – i.e. the equilibrium principle? Or do we need to find new foundations for such a field of knowledge?

In 1859 Darwin's "The Origin of Species by Means of Natural Selection" introduces a new point of view in the area of the sciences of living entities: the theory of evolution. Since then, such approach has become the ruling principle both in the theoretical and applied research activities in the above cited fields and it constitutes the foundation of the genetic engineering technologies which are one of the most advanced scientific developments of the present time. Consequently, in a gross generalization we could state that modern sciences employ two paradigms: the gravitational one, which is used in elementary physics, and the evolutionary one, concerning the sphere of the living. After all, Heraclitus got his revenge! Thus, since economics deals with human behaviour, why should it find its methodological foundation within the field of elementary physics, rather than in the realm of the living?

Let's now take into consideration the classical school of economic thought, which is based on authors such as Smith, Ricardo, Malthus, Marx and more recently Sraffa. Although each of these economists shows his own peculiarities, there is not doubt that all of them shared a common interest on how the economic structures of human societies were going to be developed and transformed. In other words, understanding the evolution of the economic phenomena was their communal goal. A task that hardly has any usage of the GEE concept of equilibrium. Besides, it is quite clear that there exists more than one point in common between the way of thinking of the classical economists and the evolutionary paradigm derived from Darwin's approach. Though generalizing about the methodology employed by the classical authors is never simple, since each of them shows unique features, some common assumptions concerning the topic under examination may be highlighted. In the classical world the forces which generate the economic phenomena may reach a temporary state of balance – a completely different concept from the GEE idea of equilibrium – or, if unbalanced, they may create the conditions for the economic structures to mutate. Both such circumstances are relative to the time and the place. The evolutionary changes that will eventually take place are not derivable from the solution of a mathematical equation, no matter how complicate it may be, although of course some aspects of the process are indeed apt to be formalized.

After we sketched the main features of the two rival paradigms – the gravitational and the evolutionary ones – we could try to answer a critical question: does the general economic equilibrium theory belong to the gravitational realm? Those who are familiar with the literature, are well aware that uniqueness and stability of the solution for the general equilibrium model have always be considered of crucial importance in order to assert the validity of that same theory. It is a well known fact that recently every hope to reach such a result has been relinquished, at least if we seek a sufficient condition which is based on microeconomic properties. On the other hand, the formal consistency of the Arrow-Debreu approach has been proven beyond any doubt. It seems that such a line of research has reached its natural limits, as far as both positive and negative results are concerned. Depending on the point of view, such a situation can be considered more or less satisfactory, even if there is no doubt that what has been obtained is somehow less than what was expected when this line of economic thought started. So we can now come to the point: can the Arrow-Debreu theory be considered a gravitational one? If we take into consideration the formalization which has been put forward by Gerard Debreu in "Theory of Value", then the answer is no, since uniqueness and stability of solutions cannot be guaranteed by conditions which show an acceptable level of generality. Still, we know by experience that we can build specialized sub-models – i.e. models based on restrictive assumptions – in such a way to obtain gravitational environments which will exist only within the chosen subset of the economic phenomena. Therefore, the Arrow-Debreu paradigm can be viewed as an abstract framework useful to theoretically support operative instruments, rather than a model in itself. More a form of specialized scientific language rather than a theory in the proper sense of the world, since the way it is structured does not allow a direct interpretation of the world phenomena. In order to achieve such a power, the Arrow-Debreu framework must be adjusted to the actual portion of the economic word we intend to analyze. From this need it derives the huge amount of specialized equilibrium models which can be found in the literature. But if equilibrium breaks down to a multiplicity of equilibria, each one rooted in the peculiar circumstances of the time, the place and the contest, how could we define such a state if not an evolutionary one? When the idea of equilibrium is brought to its maximum stage of development, it will be compelled to transmute into an evolutionary context. There is no need to add more…

In the end, let's examine the general framework underlying the process of decision making in the field of economic events. Firstly, of course, we suppose to have clearly defined the target we want to reach. In order to pursue such a scope, let's suppose that we are able to access a given set of information and to use computational tools which can be supposed up to the state of art. Clearly, it is possible that, though operating at the best of our analytical capabilities, we will be unable to define a unique strategy, but on the contrary we come up with a multiplicity of potentially optimal solutions, each one conditionally subdued to the future outcome of a number of critical events. In other words, each potential strategy may be optimal, or may be not, according to the fact that certain conditions will, or will not, be verified in the future within the economic environment where we have to operate. Since we are neither gods nor magicians, and therefore the future is obscure for us as a black cat in a dark night, we are left with a major problem or, to be more precise, a dilemma.

Dealing with it in general terms, we face three possibilities. The first one is to do nothing. In fact, we could decide that if we were unable to define a strategy which would be successful under any condition – i.e. for any possible outcome of the future critical events – it is better not to act. Even if such a choice cannot be considered an irrational one, often it is self-defeating, since it put the operator in a completely passive stance: the world will be taken "as it is", without any possibility to influence the course of the events.

The second approach is the one that we will define of a gravitational type. Since we may reasonably suppose that the past and the present can always be known if we make the effort to collect the desired records, what we can do is to extend our initially given set of information, in such a way to include any interesting item of knowledge concerning the past and the present events of the economic environment under examination. Then, using the best statistical inference tools, we can try to forecast the future values of the parameters related to the above mentioned critical events. Therefore, now we should be able to select the supposed optimal strategy among the feasible ones. The weak point of such an approach lays in the unavoidable gravitational assumption: we must suppose that the structural features of the economic world where we operate do not change in the course of time. Any entrepreneur knows well that such a conjecture is highly abstract. In economic matters the future rarely resembles the past and therefore critical choices can hardly ever be taken only according to historical records.

We are now left with the third choice: the evolutionary approach. Along with Keynes' treatment of the "animal spirits" expectations, the entrepreneur will base his investment acts upon his own subjective views relative to the future of the economic world where he is operating. Creating a subjective model of such a setting is a demanding task which implies both quantitative and qualitative aspects. As far as the quantitative estimates are concerned, we can relay on de Finetti approach based on the definition of subjective probability. In such a way we are able to subjectively measure the outcome of critical events. Still, in order to obtain a structure that can be used to forecast the evolutionary changes of the economic system, we need to model the complex structure of relationships which interconnects each element of the environment to every other one. To accomplish such a task, we can use Simon's simulacra: a computer simulation of the economic environment we are interested in. Once we possess such an instrument, and we have subjectively estimated the parameters which are needed to operate the simulation, we can bring into life our little virtual world. In other words, we can let the application run and observe how the critical variables of the model evolve through time. If the procedure has been carried on well enough, we might be able to obtain a forecast of the evolutionary changes that will take place in the economic context under examination. With regard to such a goal, we should remember Keynes' stress on "short period".

Expectations founded on "animal spirits", as well as subjective probabilistic estimates, are meaningful only on a "short run" basis, while they are useless if we pretend reliability in an asymptotic context. Such observation is supported by Simon's continuity principle which can be resumed as it follows: the more a simulation is close to the portion of reality we are analyzing, the more its evolutionary behavior will be similar to what will happen in the real world. Nevertheless, since no simulation can be identical to the simulated phenomena, the more we move further in the time scale, the more the evolution of the simulation will differ from the actual events. As a consequence of such a principle, limit patterns cannot be analyzed by using Simon's tool. Therefore, asymptotic behavior of a simulation model ought to be regarded as meaningless.

    Equilibrium? Let's get rid of such a "zombie"!